\documentclass{article}
\usepackage[legalpaper, margin=1in]{geometry}
\usepackage{authblk}

\usepackage[utf8]{inputenc}

\usepackage[space]{grffile}

\usepackage{amsmath,amssymb,wasysym}
\usepackage{subfigure}
\usepackage{siunitx}
\usepackage{caption}
\usepackage{indentfirst}
\allowdisplaybreaks

\usepackage{graphicx, dcolumn,subfigure,bm}

\usepackage{hyperref}

\title{Simple Self-calibrating Polarimeter \\ for Measuring the Stokes Parameters of Light}

\author[1,2,*]{Vitaly Wirthl}
\author[1,3]{Cristian D. Panda}
\author[1,4]{Paul W. Hess}
\author[1,5,**]{Gerald Gabrielse}
\affil[1]{\small Department of Physics, Harvard University, Cambridge, Massachusetts 02138, USA}
\affil[2]{\footnotesize Present address: Max Planck Institute of Quantum Optics (MPQ), 85748 Garching, Germany}
\affil[3]{\footnotesize Present address: University of California, Berkeley, USA}
\affil[4]{\footnotesize Present address: Department of Physics, Middlebury College, Middlebury, Vermont 05753, USA}
\affil[5]{\footnotesize Center for Fundamental Physics,				Northwestern University Department of Physics and Astronomy, Evanston, IL 60208, USA}
\affil[*]{vitaly.wirthl@mpq.mpg.de}
\affil[**]{gerald.gabrielse@northwestern.edu}

\date{}
\begin{document}
	
	\newcommand{\ta}{\tilde \alpha}
	\newcommand{\tb}{\tilde \beta}
	\newcommand{\sr}{\sqrt{2}}
	\newcommand{\cccc}{\cos(2\alpha_0-4\beta_0)}
	\newcommand{\ssss}{\sin(2\alpha_0-4\beta_0)}
	
	\newcommand{\czb}{\overline{C}_0(\ta)}
	\newcommand{\cfb}{\overline{C}_4(\ta)}
	\newcommand{\sfb}{\overline{S}_4(\ta)}

	\maketitle
	
	%\author{Vitaly Wirthl,\authormark{1,2,*} Cristian D. Panda,\authormark{1,3} Paul W. Hess,\authormark{1,4}
	%and Gerald Gabrielse\authormark{1,5,**}}
	
	%\address{\authormark{1}Department of Physics, Harvard University, Cambridge, Massachusetts 02138, USA\\
	%	\authormark{2}Present address: Max Planck Institute of Quantum Optics (MPQ), 85748 Garching, Germany\\
	%		\authormark{3}Present address: University of California, Berkeley, USA\\
	%		\authormark{4}Present address: Department of Physics, Middlebury College, Middlebury, Vermont 05753, USA\\
	%			\authormark{5} Center for Fundamental Physics,
	%				Northwestern University Department of Physics and Astronomy, Evanston, IL 60208, USA\\
	%	}

	%\email{\authormark{*}vitaly.wirthl@mpq.mpg.de} 
	%\email{\authormark{**}gerald.gabrielse@northwestern.edu} 

\begin{abstract}
A simple, self-calibrating, rotating-waveplate
polarimeter is largely insensitive to light intensity fluctuations and is shown to be useful for determining the Stokes parameters of light.   This study shows how to minimize the in situ self-calibration time, the measurement time and the measurement uncertainty. The suggested methods are applied to measurements of spatial variations in the linear and circular polarizations of laser light passing through glass plates with a laser intensity dependent birefringence.
These are crucial measurements for the ACME electron electric dipole measurements, requiring accuracies in circular and linear polarization fraction  of about $0.1\%$ and $0.4\%$, with laser intensities up to  $100\,\rm{mW}/\rm{mm}^2$ incident into the polarimeter.  
\end{abstract}

\section{Introduction}

Light polarimetry \cite{OverviewPol} is extremely important in physics \cite{OverviewPol,AtomicPhysicsPol}, plasma physics \cite{ITERPolarimeter} and astronomy  \cite{AstronomyPolar,DustAlignment,DustMagnField}.  The application that triggered this study was the ACME (Advanced Cold Molecule Electron electric dipole moment experiment) collaboration's need to understand the systematic uncertainties in electron electric dipole measurements from polarization gradients imprinted on laser beam that pass through glass under mechanical or thermal strain  \cite{Baron:2013eja, baron_methods_2017}. The simple polarimeter studied here is easy to construct with commonly available optical elements.  Incident light travels through  a linearly polarizing beam splitter followed by a quarter wave plate. Typically, one or the other of these is rotated, and the relative advantages of both options has been studied \cite{Schaefer,liang_analysis_2019}. In our polarimeter, both the linear  polarizer and the waveplate can be separately rotated about their axis. In addition to a detector that measures the intensity of the light that makes it through optical elements as a function of their rotation angles, a detector that rotates with the linear polarizer greatly reduces the sensitivity to fluctuations in the incident light intensity.  

The use of a rotating waveplate polarization analyzer is not  new \cite{Berry:77}, and the basic principles (reviewed in the Appendix) are well established.  What we sought to clarify is what calibrations and measurement methods are required in a practical device to simultaneously and accurately measure linear and circular polarization fractions in a  time-efficient way.  The precise location of the transmission axis of the linear polarizer and also the fast axis of the waveplate are typically not known initially, nor is the precise relative phase delay for the fast and slow axes of the  waveplate. This work describes the calibration parameters needed for accurate polarimetry, the minimal set of measurements that are required to do these calibrations, and the most time-efficient method for accurately measuring the polarization state of incident light.  The self-calibration methods devised and characterized using primarily a linearly polarized calibration light source are entirely in situ.  We avoid removing, inserting or reversing optical elements as is sometimes done \cite{Romerein:11}.  There is no need to modify the assembled polarimeter in any way for calibration.  We demonstrate measuring the circular polarization fraction to an accuracy of about $0.1\,\%$, and linear polarization fractions to about $0.4\,\%$.

A description of the Stokes parameters that fully characterize the information stored in light polarization is presented in Section \ref{sec:Polarimeter}, along with the concept of the simple  polarimeter that is used to measure them.  Details are reviewed and summarized in the Appendix (Sec. \ref{sec:Appendix}), along with equivalent conventions for naming the Stokes parameters.  
Section \ref{sec:extractingStokes} discusses what is needed to extract Stokes parameters from the intensity of the light that makes it through the polarimeter as the optical linear polarizer and waveplate are rotated. 
Section \ref{sec:LabRealization} describes a laboratory realization of a simple polarimeter that uses two detectors to reduce the sensitivity to fluctuations in light intensity.   Section \ref{sec:Calibration} presents  optimized, in situ calibration techniques.  An analysis of the calibration uncertainties and the resulting polarization measurement uncertainties follows in Section \ref{sec:Uncertainties}. We illustrate the usefulness of the polarimeter with an ellipticity gradient measurement in Section \ref{sec:application} and end with a conclusion in Section \ref{sec:Conclusion}. 
 
\section{Simple Polarimeter}
\label{sec:Polarimeter}

The goal is to characterize the polarization state of partially polarized light that is incident on the polarimeter.  The 4 Stokes parameters \cite{Stokes,Goldstein} (using the naming convention from \cite{Berry:77}), 
\begin{subequations}
	\label{eq:originalStokes}
	\begin{align}
		I =& ~I(0^\circ) + I(90^\circ) =  I(45^\circ) + I(-45^\circ) 		   = I_\text{RHC} + I_\text{LHC}, \\
		M =& ~I(0^\circ) - I(90^\circ),  \\
		C =& ~I(45^\circ) - I(-45^\circ), \\
		S =& ~I_\text{RHC} - I_\text{LHC},
	\end{align}
\end{subequations}
fully characterize the properties of the light.  The total intensity $I$, and the two linear polarizations $M$ and $C$, are defined in terms of intensities $I(\alpha)$ measured after the light passes through a perfect linear polarizer whose transmission axis is oriented at the indicated angles with respect to the polarization of the incoming light. The circular polarization $S$ is the difference between the intensity of right- and left-handed circularly polarized light, $I_\text{RHC}$ and $I_\text{LHC}$.  
Because they are independent of the light intensity, the normalized Stokes parameters $M/I$, $C/I$, and $S/I$  are generally more useful than the Stokes parameters themselves.
The relationships between the Stokes parameters and the electric field of the light are summarized in the appendix.  Alternatives to the labels (I,M,C,S) are  (S$_0$,S$_1$,S$_2$,S$_3$) \cite{Goldstein}, (I,Q,U,V) \cite{KligerPolLight}, and (A,B,C,D) \cite{Stokes}.

\begin{figure}[htbp!]
	\centering
	\includegraphics[width=0.7\hsize,keepaspectratio]{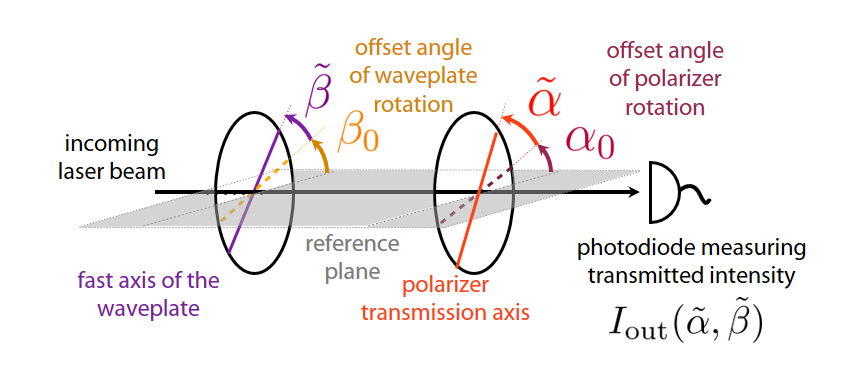}
	\caption{Polarimeter composed of a rotating waveplate  followed by a linear polarizer and a detector.  The axes of the optical elements are specified with respect to a reference plane.  }
	\label{fig:genscheme}
\end{figure}

The Stokes parameters can be determined using simple optical elements that are readily available and can be set up in most any laboratory, following Stokes \cite{Stokes} and an early experimental realization \cite{Berry:77}.  The light to be analyzed is sent through a waveplate with retardance $\delta$ followed by a coaxial linear polarizer (Fig. \ref{fig:genscheme}), along the axis perpendicular to the center of both elements.   It is crucial that both of these optical elements can be rotated in situ for calibration and to fully characterize partially polarized light that travels through the polarizer to an intensity detector. 
The fast axis of the retarder is specified by an angle $\beta = \tb +\beta_0$ with respect to a reference plane.  The angle $\tb$ is read out from an encoder attached to the retarder plate, and angle $\beta_0$ is an initially unknown offset.  Similarly, the transmission axis of the linear polarizer is specified by $\alpha = \tilde \alpha+\alpha_0$, where $\ta$ is read from an encoder attached to the polarizer, and $\alpha_0$ is an initially unknown offset angle.  Typically these mentioned axes are indicated on the optical elements.  If so, it is advantageous to assemble the polarimeter so that the magnitude of both $\alpha_0$ and $\beta_0$ are smaller than $\pi/4$.  In situ calibrations can then establish these offset angles more precisely when this is needed. The linear polarizer has an extinction ratio $r$, the factor by which the intensity of linearly polarized light is reduced when its polarization axis is perpendicular to the polarizer transmission axis.  This work describes which calibrations are needed for which measurements, how these can be done efficiently, and how these affect that accuracy of polarization measurements.  

For incident light with a Stokes vector $(I,M,C,S)$, the light intensity detected after the polarimeter varies in intensity as a function of the encoder angles $\ta$ and $\tb$, 
\begin{align}
   I_\text{out}   (\ta, \tb)   =  \tfrac{1}{2} I(1+r)   & + \tfrac{1}{2} S (1-r) \sin \delta \sin(2 \alpha_0 +2 \tilde \alpha- 2\beta_0 - 2 \tilde \beta)  \nonumber  \\   & + \,\tfrac{1}{2} C(1-r) [  \cos \delta \sin(2 \alpha_0 +2 \tilde \alpha- 2\beta_0 - 2 \tilde \beta)     \cos ( 2\beta_0+2  \tilde \beta)   \nonumber  \\ & \quad +\cos(2\alpha_0 +2 \tilde \alpha- 2\beta_0 -2  \tilde \beta)  \sin  ( 2\beta_0+2  \tilde \beta)  ] \nonumber \\   & + \,\tfrac{1}{2} M(1-r) [- \cos \delta \sin(2 \alpha_0 +2 \tilde \alpha - 2\beta_0 - 2 \tilde \beta)    \sin (2\beta_0+2  \tilde \beta) \nonumber \\ & \quad + \cos(2\alpha_0 +2 \tilde \alpha- 2\beta_0-2  \tilde \beta) \cos ( 2\beta_0+2  \tilde \beta) ]. 
\label{eq:IOut}
\end{align}
  For an ideal linear polarizer with $r=0$, this expression agrees with Stokes \cite{Stokes} for $\beta_0 = 0$ and with  \cite{Berry:77, Romerein:11}, though $\beta_0$ is defined with an opposite sign in \cite{Berry:77}.   
An appendix reviews how this expression is derived, and corrects a typo in \cite{Berry:77}.  

The detected intensity  $I_\text{out}(\ta,\tb)$  depends upon the parameters $\delta$, $r$, $\alpha_0$, and $\beta_0$, as well as upon the Stokes parameters and upon the angles $\ta$ and $\tb$ that are read out from the encoders.  These parameters, or some combinations of them, must be known to determine the incident Stokes parameters from the variations of the detected intensity as a function of $\ta$ and $\tb$. The frequency-dependent relative phase delay $\delta = \pi/2+\epsilon$ between the fast and slow axis of the waveplate is always needed. The phase is $\pi/2$ for a quarter waveplate, and $\epsilon$ is typically small.  (The optimal value of $\delta$ for statistics-limited measurements with low light levels may differ from $\pi/2$ \cite{sabatke_optimization_2000}, but we do not consider this case.)  An in situ calibration procedure for $\epsilon$ is presented.  Calibration procedures are also presented for the combinations of the small offset angles $\alpha_0$ and $\beta_0$ that are needed for some measurements, and for $\alpha_0$ and $\beta_0$ individually.  

To emphasize how the detected intensity varies as the waveplate is rotated,  
Eq. \ref{eq:IOut} can be written as a Fourier series in terms of the waveplate encoder angle $\tb$ as
\begin{align}
\label{eq:ioutFour}
I_\text{out}  (\ta,\tb) = C_0(\ta) +C_2(\ta)  \cos(2 \tilde \beta)+S_2(\ta)  \sin(2 \tilde \beta) %\nonumber \\
 +C_4(\ta)  \cos(4 \tilde \beta)+S_4(\ta)  \sin(4 \tilde \beta).
\end{align}
The corresponding Fourier coefficients for a given linear polarizer encoder angle $\ta$ are 
\begin{subequations}
\label{eq:CoefficientsFromStokes}
\begin{align}
C_0(\ta)  &=  \tfrac{1}{2} I (1+r) + \frac{(1-r) [1-\sin(\epsilon)] }{4} \left[ M \cos(2\alpha_0 + 2\tilde \alpha)  + C \sin (2\alpha_0 + 2\tilde \alpha )   \right],  \\
C_2(\ta)  &=  \tfrac{1}{2} S (1-r) \cos( \epsilon) \sin (2\alpha_0+2 \tilde \alpha-2 \beta_0) ,\\
S_2(\ta)  &= -  \tfrac{1}{2} S (1-r) \cos( \epsilon) \cos (2\alpha_0+2 \tilde \alpha-2 \beta_0) ,\\
C_4(\ta)  &= \frac{(1-r) [1+\sin (\epsilon)]}{4} [M \cos(2\alpha_0+2 \tilde \alpha-4 \beta_0) - C \sin(2\alpha_0 +2 \tilde \alpha- 4 \beta_0)]  , \\
S_4(\ta)  &=  \frac{(1-r) [1+\sin (\epsilon)]}{4} [M \sin(2\alpha_0+2 \tilde \alpha-4 \beta_0) + C \cos(2\alpha_0+2 \tilde \alpha - 4 \beta_0)].
\end{align}
\end{subequations}
These coefficients depend upon the offset angles $\alpha_0$ and $\beta_0$, and the phase delay $\delta=\pi/2 + \epsilon$ for the waveplate.  The in situ calibration of these parameters will be discussed in Sec.~\ref{sec:Calibration}.  We note (and will shortly refer to) the invariance of these equations under the simultaneous transformations $\beta_0 \rightarrow \beta_0+\pi/2$ and $S\rightarrow -S$.

\begin{figure}[htbp!]
	\centering
	\includegraphics[width=0.7\hsize,keepaspectratio]{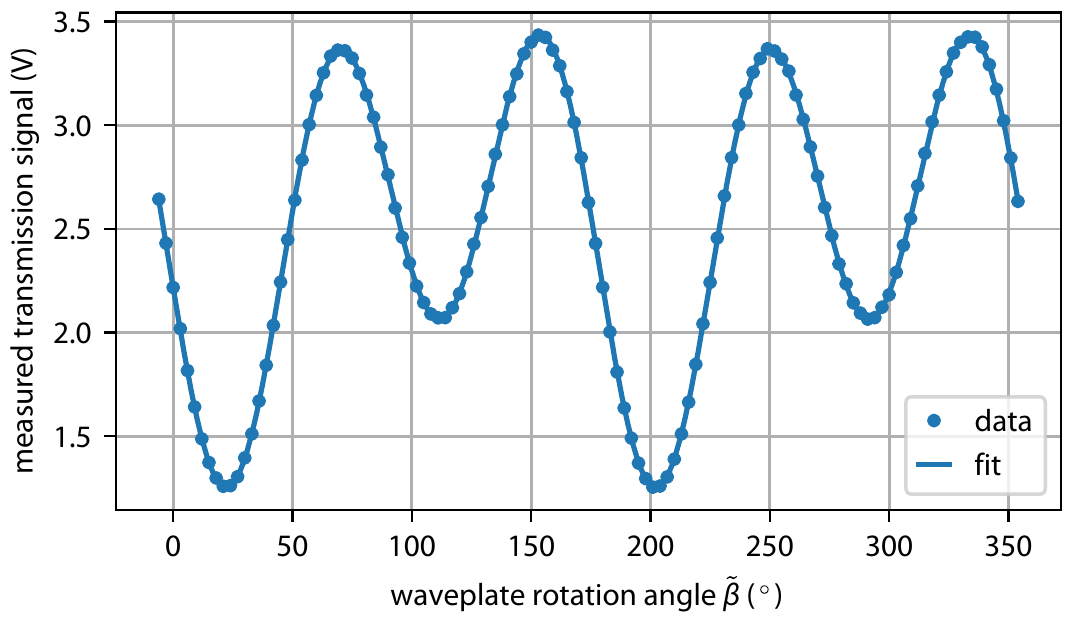}
	\caption{Illustration of how the light transmitted through the polarimeter varies with the angle of the waveplate axis as predicted by Eq. 2.}  
	\label{fig:exampleScan}
\end{figure}

Changing the waveplate encoder angle from $\tb=0$ to $\tb=2\pi$ rotates the waveplate by one complete revolution. The detected intensity after the polarimeter $I(\ta,\tb)$  varies as illustrated in Fig. \ref{fig:exampleScan}.  Components that vary as $2\tb$ and $4 \tb$ are clearly visible and the pattern repeats itself when $\tb$ changes by $\pi$ because $I_\text{out} (\ta,\tb) = I_\text{out}  (\ta,\tb+\pi)$.
For a particular choice of $\ta$, we typically determine these coefficients by Fourier transforming (or fitting) the intensity measured as the encoder angle $\tilde \beta$ makes one revolution in $3^\circ$ degree steps. In principle, fewer points are sufficient to determine the Fourier coefficients. However, in our case we are limited by the slow rotation speed of the rotation stages such that it does not take significantly more time to record more measurements for each full revolution. And, the additional points allow us to look for systematic distortions of the expected pattern.  

\section{Extracting the Stokes Parameters}
\label{sec:extractingStokes}

What can be measured is the intensity of the light after the polarimeter that is described by Eq.~\ref{eq:ioutFour} as the waveplate is rotated (as illustrated in Fig.~\ref{fig:exampleScan}).  A Fourier transform (or fit) of such variations determines the Fourier coefficients  $C_0(\ta)$, $C_2(\ta)$, $S_2(\ta)$, $C_4(\ta)$, and $S_4(\ta)$ for a particular linear polarizer orientation (given by $\ta$).  The objective is typically to determine the unknown relative Stokes parameters $M/I$, $C/I$, and $S/I$  that characterize the incident light. 

Inverting Eqs.~\ref{eq:CoefficientsFromStokes}a-e gives the Stokes parameters for the incident light as a function of the Fourier coefficients.
\begin{subequations}
\label{eq:StokesFromCoefficents}
\begin{align}
I &= \frac{2}{1+r} \left( C_0(\ta)  - \frac{1-\sin (\epsilon)}{1+\sin (\epsilon)} [C_4(\ta) \cos(4\alpha_0+4 \tilde \alpha-4 \beta_0)  + S_4(\ta)  \sin(4\alpha_0+4 \tilde \alpha - 4 \beta_0)] \right) ,\\
M &=  \frac{4}{(1-r)[1+\sin (\epsilon)]} [C_4(\ta)  \cos(2\alpha_0+2 \tilde \alpha-4 \beta_0) + S_4(\ta)  \sin(2\alpha_0+2 \tilde \alpha - 4 \beta_0)] ,\\
C &=  \frac{4}{(1-r)[1+\sin (\epsilon)]} [S_4(\ta)  \cos(2\alpha_0+2 \tilde \alpha-4 \beta_0) - C_4(\ta)  \sin(2\alpha_0 +2 \tilde \alpha- 4 \beta_0)], \\
S &= \frac{2}{(1-r) \cos(\epsilon)} \, \frac{C_2(\ta) }{\sin (2 \alpha_0+2 \tilde \alpha - 2\beta_0)} \\
&= -\frac{2}{(1-r) \cos(\epsilon)} \, \frac{S_2(\ta) }{ \cos (2 \alpha_0 +2 \tilde \alpha- 2\beta_0)} .
\end{align}
\end{subequations}
The relative Stokes parameters can thus be calculated from a set of Fourier components determined at a single linear polarizer orientation $\ta$.  Needed in addition are the offset angles $\alpha_0$ and $\beta_0$, the waveplate delay $\epsilon$, and the extinction ratio $r$, all calibrated to the needed accuracy.  

The linear polarization Stokes parameters, $M$ and $C$, are determined by $S_4(\ta) $ and $C_4(\ta)$.  No angle calibration is required to determine the intrinsically positive linear polarization intensity $L \equiv \sqrt{M^2+C^2}$ using 
\begin{equation}
 L= \frac{4}{1-r} \frac{\sqrt{ C_4(\ta)^2 + S_4(\ta)^2}}
 {1+\sin(\epsilon) }.
 \label{eq:L}
 \end{equation}
The required calibration of the waveplate delay, $\epsilon$, is discussed in Section \ref{sec:epsilon}. Distinguishing $M$ and $C$ requires in addition the cosine and sine of $2\alpha_0-4\beta_0$, the calibration of which is discussed in Section \ref{sec:MC}. Normalizing to the Stokes parameter $I$ in Eq.~\ref{eq:StokesFromCoefficents}a requires $C_0(\ta)$ in addition, as well as the cosine and the sine of $4\alpha_0-4\beta_0$, the calibration of which is discussed in Section \ref{sec:NotEssential}.  

An attractive normalization alternative determines $I$ with no knowledge of either $4\alpha_0-4\beta_0$  or $\epsilon$.  Because both $\cos(2\alpha_0 + 2\tilde \alpha)$ and $\sin(2\alpha_0 + 2\tilde \alpha) $ change sign when $\ta \rightarrow \ta + \pi/2$, the angle dependent parts in Eq.~\ref{eq:CoefficientsFromStokes}   cancel in
\begin{equation}
I =\frac{1}{1+r} [ C_0(\ta) + C_0(\ta+\pi/2)]. 
\label{eq:IfromC0}
\end{equation}
All dependence upon offset angles and waveplate delay is eliminated by measuring at two polarizer angles that differ by  $\pi/2$.

The circular polarization intensity $S$ is determined by $S_2(\ta) $ and $C_2(\ta)$. The magnitude can be determined with no knowledge of the offsets $\alpha_0$ and $\beta_0$ using  
\begin{equation}
\vert S \vert = 
\frac{2}{1-r}
\frac{ \sqrt{C_2(\ta)^2+S_2(\ta)^2}}{\vert \cos(\epsilon) \vert } 
\label{eq:SMagnitude}
\end{equation}
for any linear polarizer orientation, $\ta$. The efficient determination of the sign of S is discussed in Sec.~\ref{sec:SSign}.

For our demonstration measurements, the extinction ratio $r$  was small enough to be neglected ($r \lesssim 10^{-5}$), though this is not always true. In our case the estimated systematic uncertainty arising from finite $r$ is more than an order of magnitude smaller than other uncertainties (see Table \ref{table:uncertainties}). The contribution of $r$ is discussed in more detail in the Appendix (Sec. \ref{sec:Appendix}).

\section{Laboratory Realization with Reduced Sensitivity to Intensity Fluctuations}
\label{sec:LabRealization}

\begin{figure}[htbp!] %[htbp!]
	\subfigure[]{\includegraphics[width=0.62\hsize,keepaspectratio]{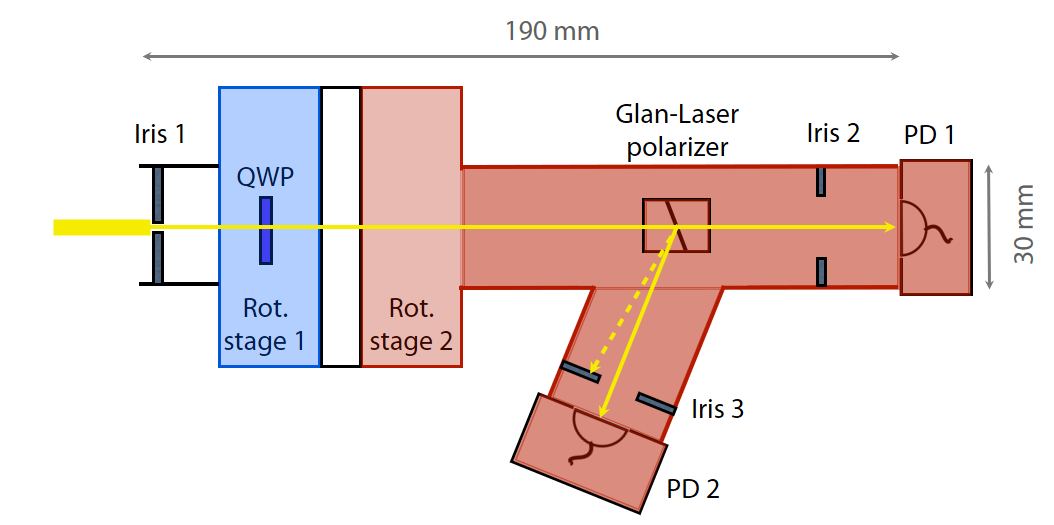}}
	\subfigure[]{\includegraphics[width=0.37\hsize,keepaspectratio]{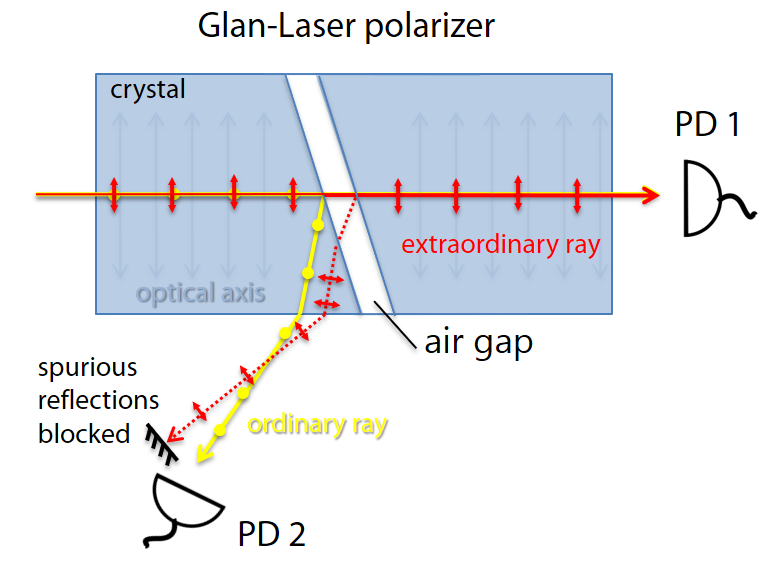}}
	\caption{(a) Scale representation of polarimeter with variable apertures, a waveplate on a rotating stage, and a linear polarizer and two detectors that rotate on a second stage. QWP: quarter waveplate, PD1 and PD2: photodetectors. (b) A Glan-laser polarizing beam splitter divides the light into transmitted and refracted beams which can be used to monitor the total intensity and correct for amplitude fluctuations in the light source. Spurious reflections of the extraordinary ray need to be blocked (only the first order reflection is drawn).}   
	\label{fig:polarScale}
\end{figure}

Our simple laboratory realization of a  rotating-waveplate polarimeter is presented to scale in Fig. \ref{fig:polarScale}(a).  The light to be analyzed enters from left through an aperture. It passes through a waveplate (labeled ``QWP'' because it is nearly a quarter wave plate) that is  mounted on a rotation stage.  The light then enters apparatus mounted on a second rotation stage.  What rotates together is a Glan-Laser polarizer that transmits one linear light polarization along the light axis, through an aperture (Iris 2) to an intensity detector PD1 (Thorlabs PDA100A).  The transverse polarization is diverted to the side of the polarizing beam splitter, through an aperture (Iris 3) to a nominally identical intensity detector (PD2) that also rotates as part of this package.   Fig. \ref{fig:polarScale}(b) is an expanded view of the polarizer.  

The apertures ensure that the light goes through the center of the optical elements.  For apertures that are too large, imperfections of optical elements (e.g. spatial inhomogeneity of the waveplate phase delay) reduce the measurement accuracy. For apertures that are too small, there are errors due to diffraction. We found that  a diameter of $1\pm0.25$ mm for the first aperture (iris 1 in Fig.~\ref{fig:polarScale}) minimized the uncertainties for our measurements at the wavelength of 1090 nm. We aligned the polarimeter with a similarly small aperture before the detector (iris 2 in Fig.~\ref{fig:polarScale}), and then opened it up during measurement to minimize diffraction errors.

Being able to independently and accurately rotate both the waveplate and the linear polarizer is a distinguishing feature of our simple polarimeter for calibration and for measuring the relative Stokes parameters $M/I$, $C/I$ and $S/I$.  The two identical precision rotation stages (Newport URS50BCC) are  specified to have a bi-directional repeatability of $0.002^\circ$ and an absolute accuracy of $0.02^\circ$.  A computer controls the stepper motors, reads the encoder angles $\ta$ for the linear polarizer and $\tb$ for the waveplate, along with the light intensity measured by both detectors as a function of these encoder angles. For a typical measurement, we rotate the  waveplate from $\ta=0$ to $\ta=2\pi$ in $3^\circ$ steps.  The orientation of the linear polarizer, given by the encoder angle $\ta$ was changed much less often, generally only between two angles, but these rotations are very important for calibrating the polarimeter and for checking for possible systematic uncertainties.   Unfortunately, for reasons not understood, our polarization measurements showed that the two rotation stages accumulated a phase error of about $0.0015^\circ$ per revolution.  Fortunately, this error was largely undone by alternating $360^\circ$ rotations in opposite directions.   
 
Fluctuations in the incident light intensity cause detection variations that can be confused with the variations caused by rotating the axes of the waveplate and linear polarizer.  However, the sum of the two detector signals should be proportional to the incident light intensity after relative gain and offset factors are determined and applied to account for detector differences. The relative gain of the detectors can be determined along with any polarization measurement by minimizing the waveplate-dependent variation of the summed signal from both detectors \cite{ThesisAndreev}.  The transmitted signal is then normalized to the sum of the deflected and transmitted light intensities. The aperture before PD2 must block the reflections of the transmitted extraordinary ray in the air gap of the Glan-Laser polarizer producing the spurious sideport beams (Fig.~\ref{fig:polarScale}(b)).

\begin{figure}[t]
	\centering
\includegraphics[width=0.7\hsize,keepaspectratio]{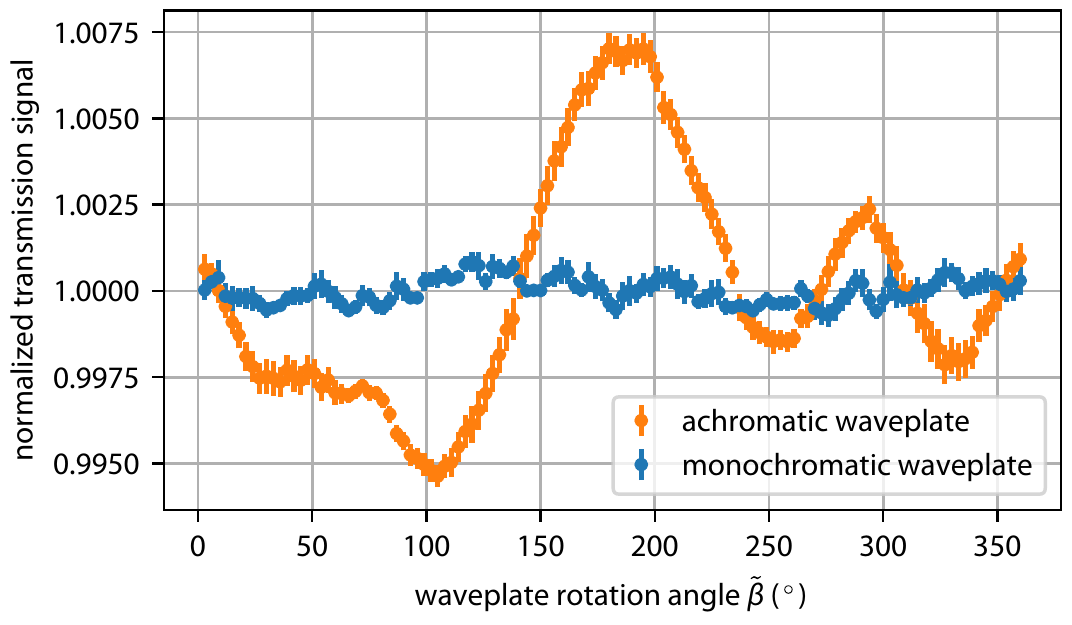}
\caption{When rotating the waveplate, the detected intensity varies significantly more for an achromatic waveplate than  a monochromatic waveplate.} \label{fig:ImperfectWaveplate}
\end{figure}

Hiqh quality optics at the wavelength of the light being analyzed is important.  For the measurements described below, a monochromatic zero-order waveplate (Thorlabs WPQ05M-1064) was used.  The fast axis is clearly marked, making it possible to easily assemble the polarimeter so that the waveplate fast axis had an offset of $\beta_0 \sim 20^\circ$ or less. There are small variations in the transmitted light intensity from an unpolarized source even after normalization, as illustrated by the blue points in Fig.~\ref{fig:ImperfectWaveplate}.  These residual variations typically contribute an uncertainty of less than $0.01\%$ in the normalized Fourier coefficients. This translates into an error in $S/I$ of smaller than $0.01 \%$.  These variations can be much worse.  The orange points in Fig.~\ref{fig:ImperfectWaveplate} show variations for an achromatic waveplate (Thorlabs AQWP05M-980).  Similar variations \cite{Harries96, Donati98} have been attributed to Fabry-Perot-type interference effects  \cite{Clarke05, Clarke04}. 

A Glan-Laser calcite polarizing beam splitter was used (Thorlabs GL10-B). This polarizer was designed to work at laser powers with a linear power density up to $2\, \text{kW}/\text{cm}$, without the damage to optical adhesives that might take place in a Wollaston prism, for example.  Its transmission axis is well marked, and it was assembled such that its transmission axis  had an offset from a desired reference axis of $\alpha_0 = 2^\circ$  or less. This polarizer was guaranteed to have an extinction ratio (discussed in the Appendix) of $r=10^{-5}$ or less which, we will see, suffices to keep the nonzero $r$ from limiting the uncertainty of our measurements.  

\begin{figure}[tb!]
	\centering
\includegraphics[width=0.7\hsize,keepaspectratio]{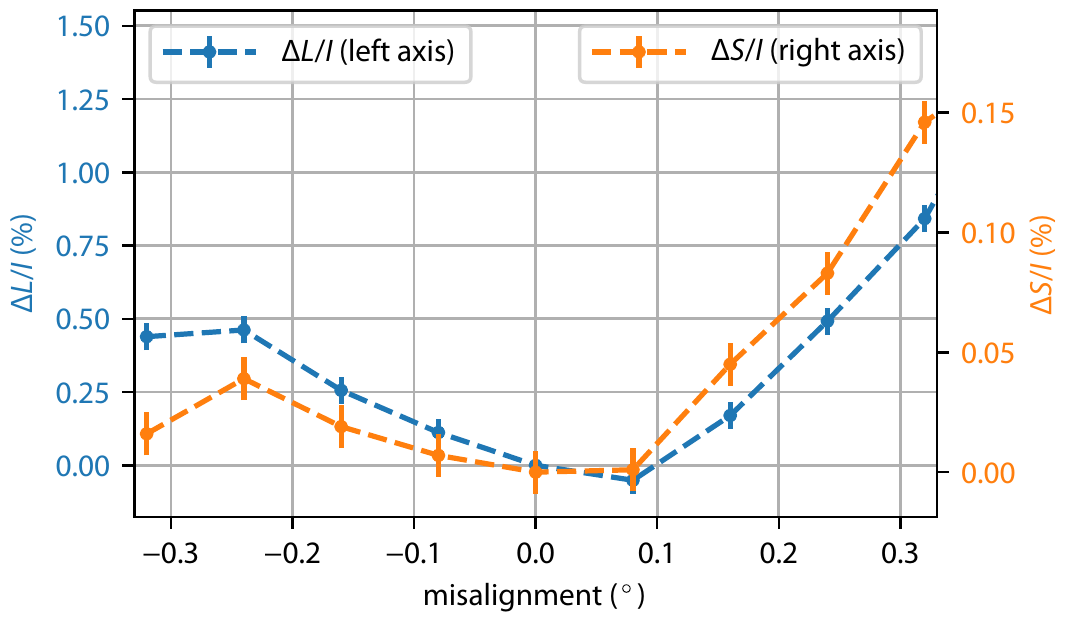}
\caption{Misalignment of the polarimeter  with respect to the light propagation axis cause systematic uncertainties in $S/I$ (orange) and $L/I$ (blue). This data was taken with a fixed incoming polarization of $S/I =16.3 \%$ and $L/I = 98.4 \% $. }
\label{fig:misalignment}
\addvspace{-5pt} 
\end{figure}

The waveplate and the polarizer that make up the polarimeter are ideally aligned so that their optical surfaces are exactly perpendicular to the direction of propagation of the laser beam. Fig.~\ref{fig:misalignment} shows an example of the systematic uncertainty that arises in measurements of $S/I$ due to a global misalignment of the polarizer with respect to the beam axis for the incident light.  We routinely align the polarimeter to better than $0.05^\circ$ which translates into uncertainties of $0.005 \%$ in $S/I$ and $0.05 \%$ in $L/I$.

Light polarization can be measured in various ways \cite{Goldstein}.  Polarimeters similar to ours, but lacking the internal calibration mechanism and immunity to  intensity fluctuations, can handle up to several $\rm{mW}/\rm{mm}^2$ \cite{HindsPol, ThorlabsPol,SKPPol} and attain uncertainties of $ {\sim}\,1\%$ in the Stokes parameters;  they have even been recommended for student labs \cite{Schaefer}.  Lower precision is also typically attained using other measurement methods.  Light is sometimes split to travel along optical paths with differing optical elements, the polarization state being deduced from the relative intensities transmitted along the paths  \cite{Azzam:92, Azzam:85, Peinado:13, He:15}. Alternatively, the light can be analyzed using optical elements whose properties vary spatially,  with the polarization revealed by the spatially varying intensity \cite{Chang:14, Zhao:14, Lepetit}.

\section{Calibration}
\label{sec:Calibration}

\subsection{Overview}

As the polarimeter is assembled, we assume that the transmission axis will be roughly aligned with the desired reference angle -- often the polarization direction of an external source of linearly polarized light.  When the encoder for the linear polarizer is set to $\ta=0$, this means that its transmission axis will be at a small offset angle $\alpha_0$ that is initially unknown. Similarly, we assumed that the fast axis of the waveplate is roughly aligned with respect to the same desired reference angle.  When the waveplate encoder is set to $\tb=0$, this means that its  fast axis is set at a small offset angle $\beta_0$, also initially unknown.  Typically, the fast axis is marked clearly so that the waveplate can be mounted in roughly the right orientation. If the fast axis has not been clearly marked, then other methods can be used to locate and mark this axis  \cite{MethodFastAxisDetermination}.  

What must be calibrated depends upon what polarization measurements are to be carried out with the polarimeter.
\begin{enumerate}

\item All polarization measurements require a calibrated delay $\delta=\pi/2+\epsilon$ between the fast and slow axis of the waveplate.  This wavelength-dependent delay can be calibrated in situ if it is not known well enough from the waveplate specification.

\item No calibration of offset angles are required to determine the Stokes parameter $I$ needed to obtain the normalized Stokes parameters $M/I$, $C/I$, $S/I$ because Eq.~\ref{eq:IfromC0} requires only measurements made at any two polarizer orientations $\ta$ that differ by $\pi/2$, 

\item If only the linear polarization $L/I$ and the  magnitude of the circular polarization $\vert S \vert$ are to be measured, no knowledge of $\alpha_0$ or $\beta_0$ is required.  

\item To measure $M/I$ and $C/I$, a calibration measurement of the angle $2 \alpha_0 - 4 \beta_0$ is needed. 

\item For small $2\alpha_0-2\beta_0$, no calibration measurement of this angle is required to get the sign of $S$ from Eq. \ref{eq:SignS}.

\item  Knowledge of the offset angle difference $\alpha_0-\beta_0$ makes it possible to determine $I$ from Eq.~\ref{eq:StokesFromCoefficents}a without changing $\ta$.

\end{enumerate}
A complete characterization of the light polarization does not require that the offset angles $2\alpha_0$ and $4\beta_0$ be known individually.  However, a way to determine these in situ is provided so these angles can be used as a diagnostic, if needed.  

\subsection{Linearly Polarized Light for Calibration}

The calibration procedure starts with a high extinction ratio polarizer placed in a light beam before it enters the polarimeter. The transmission axis of this polarizer is oriented along the desired reference direction that we wish the polarimeter to use, so that the Stokes vector for the light incident to the polarimeter is proportional to $(1,1,0,0)/2$.  The detected light intensity then simplifies to 
\begin{equation}
I_\text{out}  (\ta,\tb) = c_0(\ta) 
 +c_4(\ta)  \cos(4 \tilde \beta)+s_4(\ta)  \sin(4 \tilde \beta). 
 \label{eq:IoutLP}
 \end{equation}
The Fourier components are
\begin{subequations}
\label{eq:FourierComponentsLinearlyPolarized}
\begin{align}
c_0(\ta) =& \frac{I_c}{4}
\left[(1+r)^2 + \frac{(1-r)^2 [1-\sin(\epsilon)]}{2} \, \cos(2 \ta + 2 \alpha_0)      \right]\label{eq:C0Calibrate}\\
c_4(\ta) =& \frac{I_c}{4}
\frac{(1-r)^2 [1+\sin(\epsilon)]}{2} \, \cos(2 \ta + 2 \alpha_0  - 4 \beta_0)\label{eq:C4Calibrate}\\
s_4(\ta) =& \frac{I_c}{4}
\frac{(1-r)^2 [1+\sin(\epsilon)]}{2} \, \sin(2 \ta + 2 \alpha_0  - 4 \beta_0), \label{eq:S4Calibrate} 
\end{align}
\end{subequations}
with the constant $I_c$ giving the calibration itensity.
The factors $(1\pm r)^2$ are squared because we assume that linear polarizer used to make the linearly polarized calibration light is identical to what is in the polarimeter. Despite the square, however, the extinction ratio correction can typically be neglected if a high quality linear polarizer is used.  
The encoder angle $\ta$ for the linear polarizer is always known because it is directly read out.  The waveplate delay factor $\sin(\epsilon)$ is prominent, so linearly polarized light can be used to calibrate just the factor needed to extract the Stokes parameters for any incident light using Eqs.~\ref{eq:StokesFromCoefficents}a-e.     

Certain combinations of the offset angles $\alpha_0$ and $\beta_0$ are also present and are accessible to calibrations using linearly polarized light. In particular, the Fourier coefficient $c_0(\ta)$ depends upon $\cos(2\alpha_0)$ and $\sin(2\alpha_0)$.  The coefficients $c_4(\ta)$ and $s_4(\ta)$ depend upon $\cos(2\alpha_0-4\beta_0)$ and $\sin(2\alpha_0-4\beta_0)$.  These equations and the offset angle factors  are invariant under the symmetry invariance of the linear polarizer (under $\alpha_0\rightarrow \alpha_0+\pi$), and the invariance of the waveplate (under $\beta_0 \rightarrow \beta_0 +\pi)$. 
Because these equations are also invariant under the transformation $\beta_0 \rightarrow \beta_0+\pi/2$, the interchange of the fast and slow axes of the waveplate, the fast and slow axis cannot be distinguished from each other using only linearly polarized light sent into the polarimeter.

The $I_c$ that is proportional to the intensity  of the linearly polarized calibration light can be deduced from the Fourier coefficients in two ways, 
\begin{align}
c_0(\ta) + c_0(\ta+\pi/2) &= \tfrac{1}{2} I_c (1+r)^2 \label{eq:Norm0}\\
\sqrt{c_4(\ta)^2 + s_4(\ta)^2} &= \tfrac{1}{8} I_c (1-r)^2 [1 + \sin(\epsilon)]. \label{eq:Norm4}
\end{align}
Both are independent of the orientations of the linear polarizer and the waveplate. 
The first of these requires changing the linear polarizer orientation but is independent of the waveplate phase delay.  The second does not require changing the polarizer orientation but it does depend upon the waveplate delay.

\subsection{Waveplate Calibration with Linearly Polarized Light}
\label{sec:epsilon}

For completeness, and to emphasize the importance of a high quality linear polarizer, we explicitly display the extinction ratio $r$ in the Appendix and in the displayed equations so far.  To simplify this calibration section we assume that the extinction ratios for the linear polarizers are very small.  For both the polarimeter, and for the external calibration polarizer, we thus set $r=0$.

The linearly polarized calibration light can be used to determine the  $\sin(\epsilon)$ that contains all that must be known about the  relative phase delay between the fast and slow axes of the waveplate, $\delta=\pi/2+\epsilon$.  Eqs. \ref{eq:Norm0} - \ref{eq:Norm4} together determine
\begin{equation}
\sin(\epsilon) = \frac{4 \sqrt{ \,c_4(\ta)^2 + s_4(\ta)^2}}
{c_0(\ta) + c_0(\ta+\pi/2)}
-1.  
\label{eq:sine}
\end{equation}
There is no need to know the offsets $\alpha_0$ and $\beta_0$, no need to assume that they are small, nor must we assume that $\epsilon$ is small.  This equation does not use the measured values of $c_4$ and $s_4$ for a linear polarizer axis at $\ta+\pi/2$.  This is remedied in the more cumbersome expression \begin{equation}
\sin(\epsilon) = \frac{2\sqrt{ c_4(\ta)^2 + s_4(\ta)^2} + 2\sqrt{ c_4(\ta+\pi/2)^2 + s_4(\ta+\pi/2)^2}}
{c_0(\ta) + c_0(\ta+\pi/2)}
-1  
\label{eq:sine2}
\end{equation}
that could give a lower uncertainty
because it makes use of all of the measured values.  Choosing  $\ta=0$ allows the use of the same calibration data that determines  $2 \alpha_0$ for the offset angle calibration (see Eq.~\ref{eq:Alpha0Tan} below).  Choosing $\ta=-\pi/4$ uses calibration measurements that also  determine $2\alpha_0$ in Eq.~\ref{eq:Alpha0}.  

As an alternative, we can make use of either of two  first order expansions for small $\epsilon$ and $\alpha_0$ to determine $\epsilon$ rather than $\sin(\epsilon)$.  This first is 
\begin{equation}
  \epsilon \approx 1+ 2 \, \frac
  {  \sqrt{c_4(0)^2+s_4(0)^2}-c_0(0)}
  {\sqrt{c_4(0)^2+s_4(0)^2}+c_0(0)}. 
\label{eq:epsilon}
\end{equation}
The second is 
\begin{equation}
  \epsilon \approx \frac{1}{4}\left[      
  3 - \frac{c_0(0)}{\sqrt{c_4(0)^2+s_4(0)^2}}.
  \right]  
\end{equation}
The advantage of this approach is that the phase delay can be determined from measurements at a single linear polarizer orientation. The restriction is that this encoder angle used must be $\ta=0$. (The same set of Fourier coefficients thus cannot be used to also determine $2 \alpha_0$ using Eq. \ref{eq:Alpha0}.)  A disadvantage of the above expression is that it is the leading term of an expansion for small $\epsilon$ and $\alpha_0$.  However, we nonetheless found it useful because the corrections are very small, by factors of order $\epsilon$ and $\alpha_0^2$.  

A potentially faster possibility that we have not studied so far is  to skip the Fourier transform in favor of measuring the detected intensity $I_\text{out}(\ta,\tb)$ at 4 encoder positions, to determine the ratio
\begin{equation}
    R(\ta,\tb) =  \frac
  {-I_\text{out}(\ta,\tb)+I_\text{out}(\ta+\tfrac{\pi}{2},\tb)-I_\text{out}(\ta,\tb+\tfrac{\pi}{4})+I_\text{out}(\ta+\tfrac{\pi}{2},\tb+\tfrac{\pi}{4})}
  {I_\text{out}(\ta,\tb)+I_\text{out}(\ta+\tfrac{\pi}{2},\tb)+I_\text{out}(\ta,\tb+\tfrac{\pi}{4})+I_\text{out}(\ta+\tfrac{\pi}{2},\tb+\tfrac{\pi}{4})}.
\end{equation}
The waveplate delay is then determined by
\begin{subequations}
  \begin{align}
    \sin(\epsilon ) &= 1 +\frac{2 R(\ta,\tb) }{\cos(2\ta+2\alpha_0)}     \\
  &\approx   1 + 2 R(0,\tb).
  \end{align}
\end{subequations}
The second expression is for a choice of $\ta=0$ and  expansion for small $\alpha_0$, with corrections of order $\alpha_0^2$.  It has the attractive feature that it is independent of the angle offsets.

\subsection{Calibration of $2\alpha_0-4\beta_0$ with Linearly Polarized Light}
\label{sec:MC}

The calibration angle $2\ta -2\alpha_0-4\beta_0$ is needed to extract linear polarization components $M$ and $C$ from a measured set of Fourier components using   Eqs.~\ref{eq:StokesFromCoefficents}b-c. More precisely, since $\ta$ is an always known encoder angle, we need two parameters, $a \equiv \cos(2\alpha_0-4\beta_0)$ and $b \equiv \sin(2\alpha_0-4\beta_0)$, in terms of which Eqs.~\ref{eq:StokesFromCoefficents}b-c become
\begin{subequations}
\label{eq:MCab}
\begin{align}
 M &=  \frac{4}{1+\sin (\epsilon)} 
 \Big( 
 \cos(2\ta) \, \big[a \, C_4(\ta) + b \, S_4(\ta)  \big] + \sin(2 \ta) \, \big[ a \, S_4(\ta)  - b \, C_4(\ta)     \big]
 \Big), \hfill
\\
C &=  \frac{4}{1+\sin (\epsilon)} 
 \Big( 
 \cos(2\ta) \, \big[a \, S_4(\ta) - b \, C_4(\ta) \big] - \sin(2 \ta) \, \big[ a \, C_4(\ta) + b S_4(\ta)     \big]
 \Big), \hfill
\end{align}
\end{subequations}
Both $a$ and $b$ have the invariance of the linear polarizer (under 
$\alpha_0 \rightarrow \alpha_0+\pi$) and of the waveplate (under $\beta_0 \rightarrow \beta_0+\pi$).  In addition, both of these factors are invariant under a transformation that swaps the fast and slow axes of the waveplate, i.e.\ $\beta_0 \rightarrow \beta_0+\pi/2$.  This means that it is not necessary to distinguish these axes to measure $M$ and $C$.

With linearly polarized calibration light,
 the two needed factors can be determined, independently of the waveplate delay $\epsilon$, from Eqs.~\ref {eq:FourierComponentsLinearlyPolarized}b-c to be
 \begin{subequations}
 \label{eq:ab}
\begin{align}
    a \equiv \cos(2\alpha_0 - 4 \beta_0) &=
    \frac{c_4(\ta) \cos(2\ta) + s_4(\ta) \sin(2\ta)}
    {\sqrt{c_4(\ta)^2+s_4(\ta)^2}}\\
     b \equiv \sin(2\alpha_0 - 4 \beta_0) &=
    \frac{s_4(\ta) \cos(2\ta) - c_4(\ta) \sin(2\ta)}
    {\sqrt{c_4(\ta)^2+s_4(\ta)^2}}.
    \end{align}
\end{subequations}
Both of these factors are determined uniquely even when we do not know if it is $\beta_0$  or $\beta_0+\pi/2$ that corresponds to the fast axis of the waveplate.   By computing these factors directly, rather than using inverse trig functions to determine the angle $2\alpha_0-4\beta_0$, we avoid any restrictions on either $\alpha_0$ or $\beta_0$, or the requirement that they be small.  

A calibration measurements of $a$ and $b$ only requires one choice of $\ta$.  A simple choice, $\ta=0$, gives $a=c_4(0)/\sqrt{c_4(0)^2+s_4(0)^2}$ and $b=s_4(0)/\sqrt{c_4(0)^2+s_4(0)^2}$.  A useful (but time consuming) check on the alignment and robustness of the polarimeter is to verify that these measured factors for any particular $\ta$ agree with the values that come from averaging over a full polarizer rotation.

\subsection{Calibration Needed to Determine the Sign of $S$}
\label{sec:SSign}

The most robust way to determine the magnitude of the circular polarization has already been given in Eq.~\ref{eq:SMagnitude}.  
Determining the sign of $S$ is more difficult because the detected light is invariant under an interchange of the fast and slow axis of the waveplate, given by  $\beta_0\rightarrow \beta_0+\pi/2$ and $S\rightarrow-S$, as is evident in Eqs.~\ref{eq:CoefficientsFromStokes}a-c.  This means that a sign change for $S$ and an interchange of the fast and slow axis of the waveplate cannot be distinguished. No in situ method for distinguishing unknown locations of slow and fast axes of the waveplate is therefore available with this simple polarimeter, what is needed to determine the sign of an unknown circular polarization. We present two sign-calibration methods, one that makes use of a known fast axis for a waveplate, and another that makes use of calibration light with a known sign of circular polarization.    

If the waveplate fast axis is clearly and accurately marked, then we prefer to assemble the polarimeter so that this axis approximately lines up with the transmission axis of the linear polarizer and the desired reference axis for the polarimeter. This causes $2 \alpha_0 - 2 \beta_0$ to be small, whereupon  Eq.~\ref{eq:StokesFromCoefficents}e determines
\begin{equation}
    \text{sign}(S) = - \text{sign}[S_2(0) \cos(\epsilon)]
    \label{eq:SignS}
\end{equation}  
without the need to know either $\alpha_0$ or $\beta_0$ very precisely. 

More generally, the circular polarization $S$ is given by both of two expressions, 
\begin{align}
S &= 
    \frac{2}{\cos(\epsilon)} 
    \frac{C_2(\ta) \sin(2\ta) - S_2(2\ta) \cos(2\ta)}
    {\cos(2\alpha_0 - 2 \beta_0)} \label{eq:Sa}
    \\
   S  &= 
    \frac{2}{\cos(\epsilon)} 
    \frac{C_2(\ta) \cos(2\ta) + S_2(2\ta) \sin(2\ta)}
    {\sin(2\alpha_0 - 2 \beta_0)}.
    \label{eq:Sb}
\end{align}
A prudent choice of which equation to use avoids the complications of a vanishing denominator.   Eq.~\ref{eq:Sa} is preferred over Eq.~\ref{eq:Sb} for small $2\alpha_0-4\beta_0$, for example.
The two factors needed to determine a general $S$ from measured Fourier coefficients $C_2(\ta)$  and $S_2(\ta)$ are given in terms of the known sign of the circular polarization $s$ of the calibration light as 
\begin{subequations}
\label{eq:Angle2}
\begin{align}
\cos(2 \alpha_0 - 2 \beta_0) &=  
\text{sign}(s) \,  \cos(\epsilon)  
\, \frac{C_2(\ta) \sin(2 \ta) - S_2(2\ta) \cos(2 \ta)}
{\sqrt{C_2(\ta)^2 + S_2(\ta)^2}}
\\ 
\sin(2 \alpha_0 - 2 \beta_0) &=  \text{sign}(s) \, \cos(\epsilon)     
\, \frac{C_2(\ta) \cos(2 \ta)+   S_2(\ta) \sin(2 \ta) }
{\sqrt{C_2(\ta)^2 + S_2(\ta)^2}} . \label{eq:SinFactor}
\end{align}
\end{subequations}
If Eq.~\ref{eq:SMagnitude} is used to determine $S$, then only the signs of Eqs.~\ref{eq:Sa} - \ref{eq:SinFactor} are needed to calibrate the sign of an unknown $S$.  A good test of the polarimeter comes from comparing the factors determined as a function of the encoder angle of the polarimeter and the average over measurements that span a complete polarizer rotation.

\subsection{Useful Angles Not Strictly Needed to Measure Polarization}
\label{sec:NotEssential}

The offsets $\alpha_0$ and $\beta_0$ do not need to be individually measured to completely characterize the unknown polarization components of incident light, as we have seen.   However, individual determinations of these offset angles can be a useful diagnostic.  Also, with an accurately measured difference $2\alpha_0-2\beta_0$, the  polarimeter can be used to deduce $I$ from Eq.~\ref{eq:StokesFromCoefficents}a (rather than Eq.~\ref{eq:IfromC0}) without rotating the polarizer. 

The individual offsets can be determined using linearly polarized calibration light.  This is simpler if the polarizer is assembled to make the offsets $\alpha_0$ and $\beta_0$ small, as has been recommended. There is no need for these offsets to be exactly zero because the calibration methods presented here will allow a precise measurement of the small offsets.  

The offset $2\alpha_0$ can be determined using linearly polarized light, without making an expansion in small $\alpha_0$, $\beta_0$ or $\epsilon$.    Applying Eq.~\ref{eq:C0Calibrate} for $\ta=-\pi/4$ and $\ta=\pi/4$ gives
\begin{equation}
2\alpha_0 = \arcsin 
\left[
\frac{2}{1-\sin(\epsilon)}
\frac{ c_0(-\pi/4) - c_0(\pi/4)}  {c_0(-\pi/4) + c_0(\pi/4) } 
\right]
\label{eq:Alpha0}  
\end{equation}
which determines the sign and magnitude of a reasonably small $2\alpha_0$. We assume that $\sin(\epsilon)$ has already been determined from Eq.~\ref{eq:sine}.  

A useful alternative requires no knowledge of the waveplate phase $\epsilon$, and no expansion in small $\alpha_0$, but we do assume $2 \alpha_0$ is small enough that its tangent is in its principle value region.   This calibration also requires measurements at linear polarizer orientations of $\ta=0$ and $\ta=\pi/4$. 
\begin{subequations}
\label{eq:Alpha0Tan}
\begin{align}
2\alpha_0 &= 2\arctan\left[ 1-\sqrt{ 2   \frac{ \sqrt{s_4(0)^2+c_4(0)^2} - c_0(\pi/4)}{ \sqrt{s_4(0)^2+c_4(0)^2} -c_0(0) } }   \right]\\
&\approx 1 - 2 \frac
{   \sqrt{s_4(0)^2+c_4(0)^2} - c_0(\pi/4)   }
{    \sqrt{s_4(0)^2+c_4(0)^2} -c_0(0)     }.  
\end{align}
\end{subequations}
Eq.~\ref{eq:Alpha0Tan}b is an expansion to first order in small $\alpha_0$ that served well for the demonstration measurements.  

The cosine, sine and hence the tangent of the small angle $2\alpha_0-4\beta_0$
are given by Eqs.~\ref{eq:ab}a-b.  The arctan of this angle determines this angle over the range of $-\pi$ to $\pi$, which means that the polarimeter must be assembled such that $\vert \beta_0 \vert$ is smaller than $\pi/4$ to to make sure that $\beta_0$ represents the location of the fast axis of the waveplate.  With $2 \alpha_0$ determined separately by one of the two methods of the previous section, the result is  
\begin{equation}
4 \beta_0 = 2\alpha_0 - \arctan\left[ \frac
{s_4(\ta) \cos(2\ta) - c_4(\ta) \sin(2\ta)}
{c_4(\ta) \cos(2\ta) + s_4(\ta) \sin(2\ta)} \right] .
\label{eq:beta0}
\end{equation}
With $\alpha_0$ and $\beta_0$ separately determined, the  offset $2\alpha_0-2\beta_0$ (needed in Eqs.~\ref{eq:StokesFromCoefficents}d-e to determine the circular polarization) is also determined if it is separately known to be small. This angle can then be checked against the value determined using calibration light that is partially circularly polarized (Eq.~\ref{eq:Angle2}).

\section{Uncertainties in Relative Stokes Parameters}
\label{sec:Uncertainties}

\subsection{Overview and Statistics}

In Sec.~\ref{sec:LabRealization} we discussed the systematic uncertainties that arise from waveplate imperfections, misalignment of the incident light pointing relative to the measurement axis, and the finite extinction ratio of the polarizer. These contributions to the uncertainty in the relative Stokes parameters in our demonstration measurement are summarized later in Table~\ref{table:uncertainties}.

\begin{figure}[htbp!]
	\centering
\includegraphics[width=0.99\hsize,keepaspectratio]{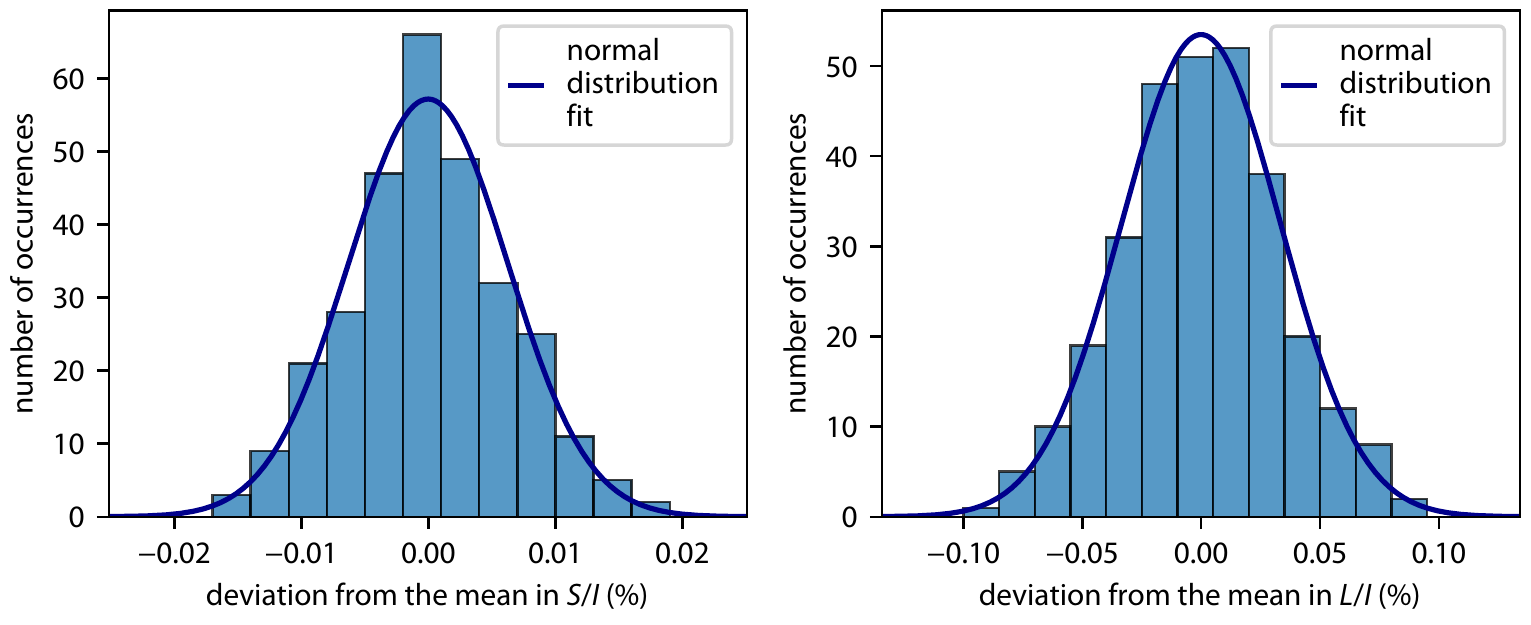}
\caption{Histograms of statistical fluctuation for 300 successive measurements of $S/I$ (left) and $L/I$ (right), along with a Gaussian fit. This example is for a large linear polarization fraction $L/I\simeq 99\%$ and a small circular polarization $S/I \simeq 3 \%$}
\label{fig:Statistics}
\addvspace{1pt} 
\end{figure}

For our demonstration measurements, the signals (e.g.\ from lasers) were large enough and our normalization procedure is robust enough that more signal averaging time no longer reduces our uncertainty.  At this point, the statistical uncertainty contribution for mostly linearly polarized light is about $\sigma_{L/I} = 0.05\%$ and $\sigma_{S/I} = 0.01\%$ (see Fig.\,\ref{fig:Statistics}).  These values are very small compared to the calibration uncertainties that are discussed next.   

\subsection{Uncertainty from the Calibration of the Waveplate $\epsilon$}
\label{sec:uncertaintiesIfromEq9}

If the polarimeter is operated with two polarizer orientations such that $I$ is determined from Eq.~\ref{eq:IfromC0}, the uncertainties in the normalized Stokes parameters ($M/I$, $C/I$ and $S/I$) that come from calibration uncertainties can be simply estimated.  Because all of the extracted Stokes parameters depend upon the relative waveplate delay phase, $\delta=\pi/2+\epsilon$, we start with the uncertainty in the Stokes parameters from the uncertainty $\sigma_\epsilon$ in $\epsilon$.  For a waveplate that is nearly a quarter wave plate, $\epsilon \approx 0$.  The example measurements make use of a waveplate for which $\epsilon \simeq 35\,\text{mrad} = 2^\circ$.  

Calibration measurements that determine $\epsilon$ are described in Eqs. \ref{eq:sine}-\ref{eq:epsilon}. The $\epsilon$ values measured in our example measurements were typically reproducible to about $0.1^\circ$.  The limit seems to be the ambient temperature variation as well as misalignments inside the polarimeter. For Eqs. \ref{eq:sine} and \ref{eq:sine2} the accuracy with which we can make a $\pi/2$ change in $\ta$ to determine $I$ is a potential source of systematic uncertainty.  The encoders of our rotation stages contribute a much smaller uncertainty than observed, as has been discussed.

The extracted $L/I$ is inversely proportional to $1+\sin(\epsilon)$.  The uncertainty $\sigma_{L/I}$ caused by an uncertainty $\sigma_\epsilon$ in $\epsilon$ for small $\epsilon$ is
\begin{equation}
\sigma_{L/I} = \frac{L}{I} \sigma_{\epsilon}.
\label{eq:LI_uncert_IfromEq9}
\end{equation}
A delay phase uncertainty $\sigma_\epsilon=0.1^\circ$ thus determines the linear polarization fraction $L/I$ with an uncertainty $\sigma_{L/I} \le 0.2\%$.  This percentage is for light that is 100\% linearly polarized; the uncertainty is proportionally smaller for lower linearly polarization fractions.  
%The blue curve in Fig. \ref{fig:UncertaintiesIfromEq9}a shows the uncertainty $\sigma_{L/I}$ as a function of $L/I$.  The same analysis and plot from waveplate phase uncertainty pertains for $M/I$ and $C/I$. 

The extracted $S/I$ is inversely proportional to $\cos(\epsilon)$.  Thus 
\begin{equation}
\sigma_{S/I} = \frac{S}{I} \epsilon \, \sigma_{\epsilon}
\label{eq:SI_uncert_IfromEq9}
\end{equation}
is smaller than for linear polarization by an additional factor of the small angle $\epsilon$, which is a factor of 30 for our example waveplate. A delay phase uncertainty $\sigma_\epsilon=0.1^\circ$ thus determines the circular polarization fraction $S/I$ with an uncertainty $\sigma_{S/I} \le 0.006\%$ for for 100\% circular polarization.  The uncertainty is proportionally smaller for lower values of $S/I$. 
%The blue curve in Fig. \ref{fig:UncertaintiesIfromEq9}b shows the uncertainty $\sigma_{S/I}$ as a function of $S/I$.  

If after calibration the polarimeter is operated with a single polarizer orientation such that $I$ is determined from Eq.~\ref{eq:StokesFromCoefficents}a, the uncertainty propagation from the calibration of $\epsilon$ is more complicated since $\epsilon$ then enters the equation for $I$ and thus all normalized Stokes parameters. We found that then the uncertainties from the calibration of $\epsilon$ to $0.1^\circ$ are larger than the above case, up to $0.35\%$ in $L/I$ and up to $0.1\%$ in $S/I$. Similar to the above consideration the uncertainties depend on the measured values (though the dependence is more complicated), e.g. for $S/I < 30\%$ the uncertainty contribution is $< 0.05\%$. Though the uncertainties are larger with the fixed polarizer operation of the polarimeter, for a given uncertainty goal this mode of operation may be advantageous due to the reduced measurement time.

\subsection{Avoiding and Minimizing Offset Angle Uncertainties}

If the polarimeter is assembled such that $2\alpha_0$ and $2\beta_0$ are small, as has been recommended, then the circular polarization fraction $S/I$  can be determined without the need to know anything more about the offset angles.  The magnitude of $\vert S\vert$ is robustly determined using Eq.~\ref{eq:SMagnitude}.  Its sign is given by Eq.~\ref{eq:SignS}. The normalization factor $I$ from  Eq.~\ref{eq:IfromC0} requires only an accurate rotation of the linear polarizer by $\pi/2$.   There is thus no contribution to $\sigma_{S/I}$ from uncertainties in the offset angles.  

The linear polarization fraction $L/I$ can be similarly determined without any knowledge of the offset angles. The $L$ can be determined from  Eq.~ \ref{eq:L}, and $I$ from Eq.~\ref{eq:IfromC0}. No knowledge of the offset angles is required, and there is thus no contribution to $\sigma_{L/I}$ from uncertainty in the offset angles.

A complete analysis of the light polarization determines $M/I$ and $C/I$ in addition to $L/I$.  Extracting these from Eqs.~\ref{eq:MCab}a-b requires knowledge of $a=\cos{\gamma }$ 
and 
$b=\sin{\gamma}$, 
where 
$\gamma=2\alpha_0-4\beta_0$.
Eqs.~\ref{eq:ab}a-b show that the uncertainties $\sigma_a$ and $\sigma_b$ should not be very different. A simple estimate of the uncertainties can be made assuming a small $\gamma$. Then 
$\sigma_b = \sigma_\gamma$ 
from Eq.~\ref{eq:ab}b. In our demonstration measurements we typically found $\sigma_b=\sigma_\gamma = 2\,\text{mrad}$, which corresponds to about 
$0.1^\circ$ 
in the offset angles.  Continuing the estimate for $\ta=0$ makes it easy to propagate the uncertainty though Eq.~\ref{eq:MCab}.  The simple estimate for the contribution to the fractional Stokes parameters from the offset angle uncertainties is $\sigma_{M/I} \approx \sigma_{C/I} \approx \sigma_b \, L/I$.  For the angle offset uncertainty mentioned and $L/I=1$, the resulting uncertainty in $M/I$ and $C/I$ is then about $0.2\%$.

\section{Application: thermally-induced birefringence}
\label{sec:application}

To illustrate the use of our internally calibrated polarimeter we measure the circular polarization induced in laser light intense enough to create thermal gradients in glass electric field plates coated with a conducting layer of indium tin oxide used in the ACME measurement of the electric dipole moment of the electron. This effect contributed to a systematic error mechanism that dominated the systematic uncertainty in a measurement that was an order of magnitude more sensitive than previous measurements \cite{Baron:2013eja,baron_methods_2017}. The polarimeter makes it possible to see whether improved electric field plates produced for a second-generation experiment succeed in reducing the thermally-induced birefringence.  The uncertainties realized for this example measurement are summarized in Table~\ref{table:uncertainties} (the polarimeter was operated with the calibrated angle difference $\alpha_0-\beta_0$ such that the Stokes vector can be normalized with a fixed polarizer orientation as described in Sec.~\ref{sec:Calibration}). 

\begin{table}[htbp!]
\centering
\caption{Summary of the systematic errors for $S/I<30\%$ and $L/I > 95 \%$.}
  \begin{tabular}{  m{4.5cm} c   c }
   \hline
  Error source  & $(L/I)_{\text{err}}$ [\%] & $(S/I)_{\text{err}}$ [\%] \\
    \hline
     ($\alpha_0 -\beta_0$) calibration to $\pm 0.1 ^\circ$& $<  0.03$ &  $< 0.005$ \\ 
    $\delta$ calibration to $\pm 0.1 ^\circ$& $< 0.35$  &  $<0.05$  \\  \hline
    Intensity normalization & $  < 0.1 $ & $  < 0.02  $ \\
     Alignment of polarimeter & $  <0.05 $ & $  <0.005  $ \\
    Imperfections of waveplate & $ < 0.012  $ &$ < 0.006$ \\
    Finite extinction ratio  & $  <0.002 $ &$ <0.002 $   \\ 
  \hline 
    Quadrature sum  & $ < 0.4$   & $ < 0.06$ \\
    \hline
  \end{tabular}
  \label{table:uncertainties}
\end{table}

To measure the polarization induced by the field plate birefringence, we start with a collimated high-power laser beam with the total power of 2~W, wavelength 1090~nm, and a circular Gaussian beam shape with waists of $w_x \simeq w_y \simeq 1.4 \, \text{mm}$. The laser beam is first polarized with the Glan-laser polarizer and then expanded in the $y$ direction using two cylindrical lenses with focal lengths of $f=10$ mm and $f=200$ mm, so that the beam shape is elongated with $w_x=1.4 \, \text{mm} \ll  w_y \simeq 30  \, \text{mm} $. The laser beam then passes through the glass plate and enters the polarimeter. $S/I$ is measured as the polarimeter is translated on a linear translation stage in the $x$ direction across the narrow illuminated area on the field plate. 

We compare the spatial gradient in $S/I$ for ACME's first- and second-generation plates.  The first-generation plate was made of borosilicate glass with a thermal expansion coefficient of $3.25 \cdot 10^{-6} \, \text{1/K}$  \cite{SchottBoro}.  The indium tin oxide layer was 200 nm thick.  The second-generation plate was designed to reduce the thermally-induced birefringence.  It is made of Corning 7980 glass with a lower thermal expansion coefficient of $0.52 \cdot 10^{-6} \, \text{1/K}$ \cite{Corning}.  To reduce absorption, the new indium tin oxide layer is thinner, at 20\,nm.

\begin{figure}[t]
	\centering
\includegraphics[width=0.8\hsize,keepaspectratio]{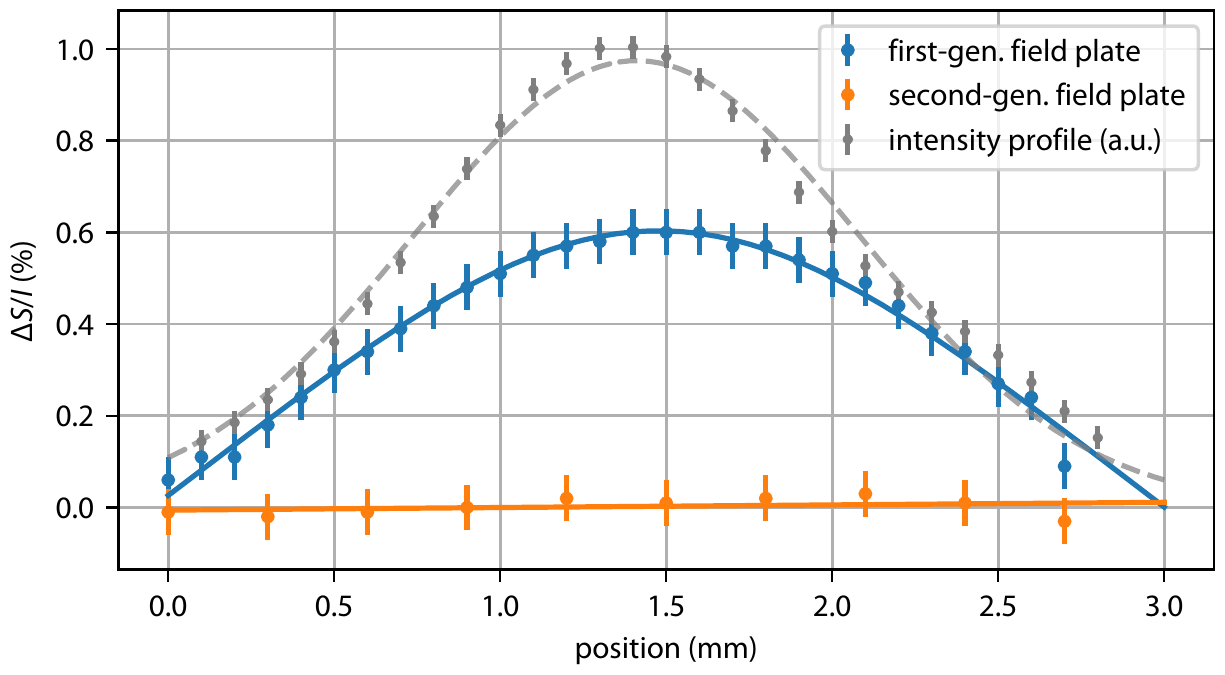}
\caption{The self-calibrated polarimeter has uncertainties low enough to compare circular polarization gradients produced by thermal gradients in first and second-generation glass field plates used by the ACME collaboration for the electron electric dipole moment experiment.  Measurements were taken with an elongated Gaussian laser beam at 1090 nm with waists $w_x=1.4 \, \text{mm} \ll  w_y \simeq 30  \, \text{mm} $ and a total power of 2 W. Error bars represent a quadrature sum of statistical and systematic uncertainties.}
\label{fig:results}
\end{figure} 

The measured changes in $S/I$ are shown in Fig. \ref{fig:results}. The intensity profile of the laser beam in the $x$ direction is the upper gray curve.  The spatial variation of $S/I$ for the first-generation plate are the blue points, with a smooth curve from a theoretical model \cite{PhDNick, PhDPaul, ThesisAndreev} that is beyond the scope of this report.  The much smaller spatial gradient of the orange points was measured with the second-generation plate.  The substantial reduction, from $S/I = 0.6\%$ to $S/I < 0.1\%$, helped to suppress the corresponding systematic error
in the ACME's second-generation measurement \cite{ACMEII}. The small uncertainties realized with the internally calibrated polarimeter are essential for this demonstration.  

Although circular polarization gradients are less than $0.1\%$ over the diameter of the laser beam, this small variation is superimposed upon a much larger $S/I \approx 8\%$ offset.  This offset can be reduced to be less than $0.1\%$ by aligning the intensity profile of the intense laser with the polarization axis.  However, the offset is a reminder that mechanical stress in optical windows and other optical elements will typically produce birefringence.  

The $8\%$ offset in our experiment comes primarily from stress in the 5.5 x 3.5 inch vacuum windows that are 0.75 inch thick, made from the same material as the field plates.  With atmospheric pressure on both sides of these windows, adjusting the tension in screws holding the windows to the vacuum chamber changed $S/I$ from about $8 \%$ to $6 \%$.  Pumping out the chamber to put a differential pressure of one atmosphere across such a window typically changed $S/I$ by up to $3 \%$.   Related measurements with the polarimeter showed that optical elements such as a zero-order half-waveplate could produce residual circular polarization of up to $3 \%$, which shows that such imperfections of linear polarized light are common. We did not observe unexpected linear polarization changes from the windows larger than the systematic uncertainties in the measurement.

\section{Conclusion}
\label{sec:Conclusion}

Optimized measurement and calibration methods are developed and demonstrated for measuring the unknown linear and circular polarization fractions of light entering an easy-to-construct and easy-to-operate polarimeter. The polarimeter has a relatively high power handling capability, a substantial immunity to fluctuations in light intensity, and it can be calibrated internally and in situ, without the need for removing or realigning any optical elements. In a demonstration measurement the circular polarization fraction $S/I$ is measured with an uncertainty below $0.1 \%$, while the linear polarization fraction $L/I$ uncertainty is below $0.4 \%$. This determines the polarization gradients due to thermally-induced birefringence in a glass field plate that were an important source of systematic error in the most precise measurement of the electron electric dipole moment.

\section{Appendix}
\label{sec:Appendix}

\subsection{Partially and Fully Polarized Light}
Elliptical polarization is the most general state of a fully polarized plane wave traveling in the $z$ direction with frequency $\omega$ and wavenumber $k$.  In cartesian coordinates, the electric field is 
\begin{equation}
\label{eq:EfieldPlaneWave}
   \bm{\vec{ \mathcal{E}}} =  \hat{\bf x}\,  \mathcal{E}_{0x}\cos(\omega t - kz+\phi)  + \hat{\bf y} \,\mathcal{E}_{0y}\cos(\omega t - kz ), 
\end{equation}
where the real $ \mathcal{E}_{0x}$ and $ \mathcal{E}_{0y}$ give the strength of each of these orthogonal electric field components. The angle $\phi$ is the phase difference between the two orthogonal components.   

The Stokes vector \cite{Stokes} defined in Eq.~\ref{eq:originalStokes} provides a useful alternative way to specify the general state of an elliptically polarized light.  They also make it possible to easily generalize to the case of partially polarized and even unpolarized light, in which case the total intensity of the light going into a detector $I = I_u + I_p$, where $I_u$ is the intensity of the unpolarized components of the light, and $I_p$ is the intensity of the elliptically polarized light, $I_p = \mathcal{E}_{0x}^2 +  \mathcal{E}_{0y}^2$. The Stokes vector  for light traveling in the $\hat{z}$ direction is    
\begin{equation}
\vec{S}=\begin{pmatrix} I \\ M \\ C \\ S \end{pmatrix} =\begin{pmatrix}   I_u +  \mathcal{E}_{0x}^2 +  \mathcal{E}_{0y}^2  \\  \mathcal{E}_{0x}^2  -   \mathcal{E}_{0y}^2 \\ 2 \,  \mathcal{E}_{0x}  \mathcal{E}_{0y} \cos \phi   \\  2 \,  \mathcal{E}_{0x} \mathcal{E}_{0y} \sin \phi    \end{pmatrix}. 
\label{eq:Stokes}
\end{equation} 
The $M$ and $C$ values quantify the light's linear polarization with respect to two sets of axes rotated by 45 degrees.  $S$ is the circular polarization of the light.  Often the polarization of the light is characterized by the linear polarization fractions, $M/I$ and $C/I$, along with the circular polarization fraction, $S/I$.

The square of the polarized intensity
\begin{equation}
I_p^2 = M^2+C^2+S^2
\label{eq:imcs}
\end{equation}
is the sum in quadrature of the three other Stokes parameter. 
The linear polarization fraction is  \begin{equation}
 L/I_p= \sqrt{(M/I_p)^2+(C/I_p)^2}   
\end{equation} 
and the circular polarization fraction is $S/I_p$. Thus  
\begin{equation}
(L/I_p)^2+(S/I_p)^2 =1.
\label{eq:LiSi}
\end{equation}
is a restatement of Eq.~\ref{eq:imcs}.  Because the relative intensities are summed in quadrature, a plane wave with  nearly complete linear polarization  (e.g.\ $L/I_p = 99 \%$) corresponds to a circular polarization that is still substantial (e.g.\ $S/I_p = 14 \%$).

\subsection{Detected Intensity}

The intensity of the light at the polarimeter's detector is given by Eq.~\ref{eq:IOut}.  The incident  Stokes vector, $\vec{S} = (I,M,C,S)$ is transformed to $B(\beta) \, \vec{S}$ as the light goes through the waveplate.  The Mueller matrix $B(\beta)$ for a waveplate whose fast axis is oriented at an angle $\beta$ with respect to a reference plane, and whose orthogonal slow axis delays the light transmission by an angle $\delta$ is \cite{Goldstein, ClarkePolLight, KligerPolLight, HandbookOptics, Shurcliff} 
%\begin{widetext}
\begin{equation}
	B(\beta) = \begin{pmatrix}  1 & 0 & 0 & 0 \\ 0 & \cos^2 2 \beta + \cos \delta \sin^2 2 \beta & \cos 2\beta \sin 2\beta (1- \cos \delta) & - \sin 2 \beta \sin \delta \\ 0 & \cos 2 \beta \sin 2 \beta (1- \cos \delta) & \cos{\delta} \cos^2 2 \beta + \sin^2 2 \beta & \cos 2 \beta \sin{\delta} \\ 0 & 
		\sin 2 \beta \sin \delta & - \cos 2 \beta \sin{\delta} & \cos{\delta}    \end{pmatrix}.
\end{equation}
Because  $B(\beta+\pi)=B(\beta)$, the waveplate for angles $\beta = 0$ to $\pi$ has the same optical properties as for $\beta = \pi$ to $2 \pi$. This matrix can be derived by inserting the electric fields after the waveplate  Eqs.~\ref{eq:Stokes}.  Unpolarized light with $I\ne0$ and $M=C=S=0$ emerges as unpolarized light with the same intensity $I$.  The waveplate transforms $M$, $C$ and $S$ between themselves.

A Stokes vector $\vec{S}$ is transformed to $A(\alpha,r)~\vec{S}$ after it passes through a linear polarizer.  The Mueller matrix for a linear polarizer \cite{Goldstein, ClarkePolLight, KligerPolLight, HandbookOptics, Shurcliff} with its transmission axis oriented at an angle $\alpha$ with respect to the reference plane is 
\begin{equation}
A(\alpha,r)  =  \begin{pmatrix} 
\frac{1+r}{2} & \frac{1-r}{2} \cos(2\alpha) & \frac{1-r}{2} \sin(2 \alpha) & 0 
\\ \frac{1-r}{2} \cos(2\alpha) 
& \frac{(1+\sqrt{r})^2}{4} + \frac{(1-\sqrt{r})^2}{4} \cos(4\alpha) 
& \frac{(1-\sqrt{r})^2}{4} \sin(4\alpha) & 0
\\ 
\frac{1-r}{2} \sin(2 \alpha) 
&  \frac{(1-\sqrt{r})^2}{4} \sin(4\alpha) 
& \frac{(1+\sqrt{r})^2}{4} -\frac{ (1-\sqrt{r})^2}{4} \cos(4\alpha)  
& 0 
\\ 0 & 0 & 0 & \sqrt{r}   \end{pmatrix}.
\end{equation}
An ideal linear polarizer has $r=0$.  A simple model of an imperfect  polarizer is incorporated in this Mueller matrix \cite{HandbookOptics}, namely that light is transmitted perfectly along the polarizer's transmission axis,  the transmitted electric field along an orthogonal axis is reduced by a factor of  the square root of the extinction ratio $r$.  An ideal polarizer has $r=0$, but available polarizers have extinction ratios that range from $10^{-2}$ to $10^{-6}$. We will include this factor in the equations of this paper to emphasize the important of using a high quality polarizer, but will generally assume that we can neglect $r$ with respect to $1$ by using a good linear polarizer with a high extinction ratio.  The polarizer used in our example experiment has $r \lesssim 10^{-5}$.  The optical properties for orientations $\alpha=0$ to $\pi$ repeat for orientations $\alpha = \pi$ to $2 \pi$ so  $A(\alpha+\pi)=A(\alpha)$.  
This matrix can be derived by inserting the electric fields after a linear polarizer in  Eqs.~\ref{eq:Stokes}.

A linear polarizer transforms unpolarized light with a Stokes vector $(I_0,0,0,0)$ into partially linearly polarized light with Stokes vector $\tfrac{1}{2}I_0\,[1+r,(1-r)\, \cos{2\alpha},(1-r)\,\sin{2\alpha},0]$.  If the transmission axis of the linear polarizer is taken to be the desired reference axis, as we suggest for calibration, then the Stokes vector for the calibration light is $\tfrac{1}{2}I_0\,[1+r,1-r,0,0]$ with no circular polarization. However, typically stress-induced birefringence, which has been neglected here, introduces a small residual circular polarization fraction.  To illustrate the robustness of such production of linearly polarized calibration light, in the worst case that circularly polarized light is substituted for the unpolarized incident light the calibration light acquires only a circular polarization intensity $S=I_0 \sqrt{r}$.  For our polarizer, this is a very small contamination of the calibration light, given than $\sqrt{r} \approx 10^{-3}$,

Returning to the general case, the intensity of the  light arriving at the detector is the intensity component of the Stokes vector for this light.  
To include the angle offsets in Fig.~\ref{fig:genscheme},  we let  $\alpha = \tilde \alpha + \alpha_0$ and $\beta = \tilde \beta + \beta_0$ in the Mueller matrices.  The detected intensity is then calculated from
\begin{equation}
\vec{S}_\text{out}(\tilde \alpha,\tilde \beta,\delta) = A(\tilde \alpha+\alpha_0,r) \cdot B(\tilde \beta+\beta_0,\delta)~\vec{S}.   
\end{equation}
Eq.~\ref{eq:IOut} results from the first component of the above outgoing Stokes vector $\vec{S}_\text{out}$ when the matrices are multiplied and the trigonometric functions are simplified. The corresponding expression in  Eq.\,(16) of \cite{Berry:77} must be fixed by replacing $4 \beta_0$ by $2\beta_0$, and $\sin^2 \delta$ with $\vert \sin(\delta) \vert$.

\section*{Funding}
U.S. National Science Foundation (NSF), German Academic Exchange Service (DAAD).

\section*{Acknowledgement}
We thank our colleagues,  D. Ang, D. DeMille, J. M. Doyle, N. R. Hutzler, Z. Lasner, B. R. O'Leary,  A. D. West, E. P. West, who worked with us to understand the thermally-induced birefringence in the ACME field plates and to design the improved second-generation field plates. They also made useful comments on the manuscript.

\section*{Disclosures}
The authors declare no conflicts of interest.

\bibliographystyle{ieeetr}

\bibliography{polarimeterPaper}

\end{document}